\documentclass[proceedings]{JHEP} 

\usepackage{epsfig}                     

\newbox\mybox
\newcommand\fverb{\setbox\mybox=\hbox\bgroup\verb}
\newcommand\fverbdo{\egroup\medskip\noindent\fbox{\unhbox\mybox}\ }
\newcommand\fverbit{\egroup\item[\fbox{\unhbox\mybox}]}

\font\beeg=cmr17 scaled 1600            
\newcommand\init[1]{\setbox\mybox=\hbox{{\beeg #1}~}%
                   \noindent\global\hangindent=\wd\mybox\global\hangafter-2%
                   \sc\smash{\llap {\lower 13.2pt \box\mybox}}}

\title{Effective theories and hadronic bound states
\footnote{Review of recent work done in collaboration with
A.~Gall, J.~Gasser, M.A.~Ivanov, E.~Lipartia, and V.E.~Lyu\-bo\-vit\-skij.}}

\author{A.\ Rusetsky\\
ITP, University of Bern, Sidlerstrasse 5, 3012 Bern, Switzerland, and\\
HEPI, Tbilisi State University, University st. 9, 380086, Tbilisi, Georgia\\
Email: \email{rusetsky@itp.unibe.ch}}

\conference{Heavy Quark Physics 5, Dubna, Russia, 6-8 April 2000}

\abstract{Recent progress in the description of the 
properties of hadronic atoms on the basis of non-relativistic effective
Lagrangian approach and Chiral Perturbation Theory (ChPT) is reported.
For the case of the $\pi^+\pi^-$ atom decay, the problem is completely 
solved, both conceptually and numerically. For the $\pi^- p$ atom, 
a general expression for the ground-state energy is obtained in the first 
non-leading order in isospin breaking, and the numerical analysis is carried 
out at $O(p^2)$ in ChPT. We briefly consider a possible solution of the 
``potential model puzzle'' in the hadronic atom problem, providing 
a constructive algorithm for the derivation of the isospin-breaking part of 
the short-range hadronic potential from field theory.}

\begin{document} 

Below, we shall report on the recent progress achieved in the description of
the hadronic bound states - so-called hadronic atoms - within the framework
of field theory. The reason why such a report is delivered at the workshop on 
heavy quark physics, lies in a close similarity of the methods that are
employed in these, otherwise very distinct fields. In fact, it turns out that
the non-relativistic effective Lagrangian technique that was proposed
originally by Caswell and Le\-pa\-ge~\cite{Lepage} 
to study QED bound states in
general, and that is widely used now in the physics of heavy quarks, provides
the most elegant and economical approach to the solution of hadronic atom
problem. In a latter case, one is faced with a theoretical challenge of
producing a merger of the effective non-relativistic theory that implies
the expansion in the inverse masses of hadrons (including Goldstone bosons)
and, therefore, in the inverse powers of the quark masses, and the ChPT
that is based on the expansion in quark masses. Such a merger can indeed be
worked out, and the systematic chiral expansion of the observables of the 
hadronic atoms can be ensured in this theory.

Recent years have seen a  growing interest in the study of hadronic atoms. 
At CERN, the DIRAC collaboration~\cite{DIRAC} aims to measure the 
$\pi^+\pi^-$ atom lifetime to $10\%$ accuracy. This would allow one to 
determine the difference $a_0-a_2$ of $\pi\pi$ scattering lengths with $5\%$
precision. This measurement provides a crucial test for the large vs small
condensate scenario in QCD: should it turn out that the quantity $a_0-a_2$ 
is different from the value predicted in standard ChPT~\cite{ChPT}, one has 
to conclude~\cite{Stern} that spontaneous  chiral symmetry breaking in QCD 
proceeds differently from the widely accepted picture. In the experiment 
performed at PSI~\cite{PSI1,PSI2}, one has measured the strong energy-level 
shift and the total decay width of the $1s$ state of pionic hydrogen, as 
well as the $1s$ shift of pionic deuterium. These measurements  yield 
isospin symmetric $\pi N$ scattering lengths to an accuracy which is unique 
for hadron physics. A new experiment on  pionic hydrogen at PSI has recently
been  approved. It will allow one to measure the decay 
$A_{\pi^-p}\rightarrow \pi^0n$  to much higher accuracy and thus enable one, 
in principle, to determine the $\pi N$ scattering lengths from  data on 
pionic hydrogen alone. This might vastly reduce the model-dependent 
uncertainties that come from the analysis of the three-body problem in
$A_{\pi^-d}$. Finally, the DEAR collaboration~\cite{DEAR} at the DA$\Phi$NE 
facility plans to measure the energy level shift and lifetime of the $1s$ 
state in $K^{-}p$ and $K^-d$ atoms - with considerably higher precision 
than in the previous experiment carried out at KEK~\cite{KEK} for
$K^-p$ atoms. It is expected~\cite{DEAR} that this will result in a 
precise determination of the $I=0,1$ $S$-wave  scattering lengths.
It will be a challenge for theorists to extract from this new information 
on the $\bar{K}N$ amplitude at threshold a more precise value of e.g. the 
isoscalar kaon-sigma term and of the strangeness content of the nucleon.

In order to carry out the precision experimental tests of QCD mentioned 
above, on the theoretical side one faces the problem of finding the
relation between the measured characteristics of hadronic atoms - energy
levels and decay probabilities - and the strong hadronic scattering lengths
in the isospin limit. In general, we have the following relations between
these quantities
\begin{eqnarray}\label{deser}
&&\Delta E_{{\rm str}}\sim
\Psi_0^2\, {\rm Re}\, a_{cc}\,(1+\delta_\epsilon)\, ,
\\[2mm]
&&\Gamma_{c0}\sim({\rm phase~space})\times \Psi_0^2\, 
|a_{c0}(1+\delta_\Gamma)|^2\, .
\nonumber
\end{eqnarray}
Here $\Delta E_{{\rm str}}$ denotes the strong energy-level shift in the $1s$
state from its Coulomb value 
(total shift minus the pure QED contribution), and $\Gamma_{c0}$ is
the partial decay width into the neutral channel
(e.g. $A_{\pi^-p}\rightarrow \pi^0n$); $a_{cc}$ and $a_{c0}$ stand for the
particular isospin combinations of the strong scattering lengths
(charged and neutral channels are marked by subscripts ``c'' and ``0'',
respectively), and $\Psi_0$ denotes the value of the Coulomb wave function
at the origin. The quantities $\delta_\epsilon$ and $\delta_\Gamma$ stand
for the isospin-breaking corrections and, in general, depend on the details 
of the strong dynamics. In order to extract the scattering lengths from the
experiments, these quantities must be known to an accuracy that matches the
accuracy of the measurements.

Historically, the potential model was the first one been applied to the
calculation of the had\-ro\-nic atom characteristics. In brief, this model 
assumes that the Coulomb effects and the mass differences between the charged
and neutral particles in the same multiplet are responsible for all
isospin-breaking effects in the low-energy hadron physics: the short-range
potential that describes strong interactions, is presumed to be
isospin-symmetric. In this manner, one may calculate the isospin breaking
corrections $\delta_\epsilon$ and $\delta_\Gamma$~\cite{PSI1,Rasche}. The
results of these calculations are however in the striking disagreement with
the field-theoretical evaluations based on ChPT (see below). It turns out
that the assumption about the isospin symmetry of the short-range hadronic
potential is too restrictive - one should allow for a small isospin-breaking
piece in the short-range interactions as well, in order to achieve an
agreement with the field-theoretical calculations. In this way - providing a
constructive algorithm for the derivation of the isospin-breaking piece from
field theory - one may achieve the solution of apparent potential model
puzzle.

Recently, using a  non-relativistic effective Lagrangian framework,
a general expression for the decay width $\Gamma_{A_{2\pi}\to\pi^0\pi^0}$
of the $1s$ state of the $\pi^+\pi^-$ atom  was obtained at 
next-to-leading order in isospin-breaking~\cite{Bern1}. Numerical analysis
of this quantity was carried out at order $O(e^2p^2)$ in ChPT~\cite{Bern2}.
These investigations have confirmed and generalized  the results of earlier 
studies~\cite{Sazdjian,Dubna}. The expression for the decay width has the 
form
\begin{eqnarray}\label{general-pipi}
&&\Gamma_{A_{2\pi}\to\pi^0\pi^0}=\frac{2}{9}\,\alpha^3 p^\star 
{\cal A}_{\pi\pi}^{~2} (1+K_{\pi\pi})\, ,
\nonumber\\[2mm]
&&{\cal A}_{\pi\pi}=a_0-a_2+O(\alpha,(m_d-m_u)^2)\, , 
\nonumber\\[2mm]
&&K_{\pi\pi}=
\frac{\Delta M_\pi^2}{9M^2_{\pi^+}}\,(a_0+2a_2)^2
-\frac{2\alpha}{3}\,(\ln\alpha-1)\times
\nonumber\\[2mm]
&\times&(2a_0+a_2)+o(\alpha,(m_d-m_u)^2)\, . 
\end{eqnarray}
Here $p^\star=(M_{\pi^+}^2-M_{\pi^0}^2-\frac{1}{4}M_{\pi^+}^2\alpha^2)^{1/2}$,
 and $a_I$, $(I=0,2)$ denote the strong $\pi\pi$ scattering lengths
in the channel with total isospin $I$. The quantity ${\cal A}_{\pi\pi}$ is
calculated as follows~\cite{Bern1}. One calculates the relativistic amplitude
for the process $\pi^+\pi^-\rightarrow\pi^0\pi^0$ at $O(\alpha,(m_d-m_u)^2)$
in the normalization chosen so that at $O(1)$ the amplitude at threshold 
coincides with the difference $a_0-a_2$ of (dimensionless) 
$S$-wave $\pi\pi$ scattering lengths. Due to the presence of virtual photons,
the amplitude is multiplied by an overall Coulomb phase $\theta_c$ 
that is removed. The real part of the remainder contains  terms
that diverge like $|{\bf p}|^{-1}$ and $\ln 2|{\bf p}|/M_{\pi^+}$ at
$|{\bf p}|\rightarrow 0$ (${\bf p}$ denotes the relative 3-momentum
of charged pion pairs). The quantity ${\cal A}_{\pi\pi}$
 is obtained by subtracting these divergent pieces, and by then 
evaluating the remainder at ${\bf p}=0$.
\begin{eqnarray}\label{threshold-pipi}
{\rm Re}\,({\rm e}^{-i\theta_c}\, t_{\pi\pi})&\rightarrow&
\frac{b_1}{|{\bf p}|}+b_2\ln\frac{2|{\bf p}|}{M_{\pi^+}}+
\frac{8\pi}{3M_{\pi^+}^2}\,{\cal A}_{\pi\pi}
\nonumber\\[2mm]
&&
\end{eqnarray}

As it is seen explicitly from Eq.~(\ref{general-pipi}), one can directly 
extract the value of ${\cal A}_{\pi\pi}$ from the measurement of the  
decay width, because the correction  $K_{\pi\pi}$ is very small and the 
error introduced by it is negligible. We emphasize that in derivation of 
Eq.~(\ref{general-pipi}), chiral expansions have not been used.
On the other hand, if one further aims to extract strong scattering
lengths from  data, one may invoke ChPT
and to relate the quantities ${\cal A}_{\pi\pi}$ and
$a_0-a_2$ order by order in the chiral expansion. This requires
the evaluation of isospin-breaking corrections to the scattering amplitude.
At order $O(e^2p^2)$ in chiral expansion we obtain~\cite{Bern2}
(for the values of scattering lengths $a_0=0.206$, $a_2=-0.0443$)
\begin{eqnarray}\label{numerics}
&&A_{\pi\pi}=a_0-a_2+\epsilon\, ,\quad
\epsilon=(0.58\pm 0.16)\cdot 10^{-2}\, ,
\nonumber\\[2mm]
&&K=1.07\cdot 10^{-2}\, ,\quad\quad
\delta_\Gamma=0.058\, .
\end{eqnarray}
At this stage, one may recall that the calculations in the potential model
with the isospin-symmetric strong potential yields $\delta_\Gamma$ being
of negative sign and the same order of magnitude~\cite{Rasche}.

In the case of the $\pi^- p$ atom, the treatment proceeds along the lines
very similar to those for $\pi^+\pi^-$ case~\cite{Bern3}. Our investigations
are aimed at the derivation of the general expression for the $\pi^- p$
atom energy-level shift in the $1s$ state. The total shift is given by a sum
of the electromagnetic (pure QED) and strong pieces. Our calculations for
the electromagnetic shift within a high accuracy yield the same result as given
in Ref.~\cite{PSI1}. The final result for the strong shift in the first
non-leading order in isospin breaking is given in a form similar to
Eq.~(\ref{general-pipi})
\begin{eqnarray}\label{general-piN}
&&\Delta E_{{\rm str}}=-2\alpha^3\mu_c^2\, {\cal A}_{\pi N}\,(1+K_{\pi N})
\\[2mm]
&&K_{\pi N}=2\alpha\mu_c(1-\ln\alpha){\cal A}_{\pi N}+o(\alpha,m_d-m_u)\, ,
\nonumber
\end{eqnarray}
where $\mu_c$ denotes the reduced mass of the $\pi^- p$ pair, and the quantity
${\cal A}_{\pi N}$ is defined analogously to ${\cal A}_{\pi\pi}$. To calculate
this quantity, one has to evaluate the $\pi^- p\rightarrow\pi^- p$ 
relativistic scattering amplitude at $O(\alpha,m_d-m_u)$, drop all 
diagrams that are made disconnected by cutting one photon line, and
discard the spin-flip piece. The remainder
is denoted by $\bar t_{\pi N}$. The regular part of $\bar t_{\pi N}$ at
threshold defines the quantity ${\cal A}_{\pi N}$ in analogy to
Eq.~(\ref{threshold-pipi})
\begin{eqnarray}\label{threshold-piN}
{\rm Re}\,({\rm e}^{-2i\theta_c}\, \bar t_{\pi N})\rightarrow
\frac{B_1}{|{\bf p}|}+B_2\ln\frac{|{\bf p}|}{\mu_c}-
\frac{2\pi}{\mu_c}\,{\cal A}_{\pi N},
\nonumber\\
\end{eqnarray}
and the normalization of the amplitude 
is chosen so that ${\cal A}_{\pi N}=b_0-b_1+O(\alpha,m_d-m_u)$, where
$b_0$ and $b_1$ denote the isospin even and odd strong $\pi N$ scattering
lengths.
  
In order to extract the value of $b_0-b_1$ from the $\pi^- p$ measurement, one
may again resort to ChPT, to calculate the isospin-breaking corrections
to the $\pi N$ scattering amplitude at threshold. At chiral order $O(p^2)$
where only the tree diagrams contribute, the result looks as follows 
\begin{eqnarray}\label{amplitude}
&&{\cal A}_{\pi N}^{(2)}=b_0-b_1+\epsilon_{\pi N}^{(2)}
\nonumber\\[2mm]
&&\epsilon_{\pi N}^{(2)}=\frac{m_p(8c_1\Delta_\pi-4e^2f_1-e^2f_2)}
{8\pi(m_p+M_{\pi^+})F^2}\, , 
\end{eqnarray}
where $\Delta_\pi=M_{\pi^+}^2-M_{\pi^0}^2$ denotes the pion mass difference,
and $c_i$ ($f_i$) are the strong (electromagnetic) low-energy constants (LEC's) 
from the $O(p^2)$ Lagrangian of ChPT~\cite{Lagrangian}.
In order to perform the numerical analysis,
one has to specify the values of these LEC's. The "strong" constant
$c_1$ can be determined from the fit of the elastic $\pi N$ scattering
amplitude at threshold to KA86 data~\cite{private}: 
$c_1=-0.925~{\rm GeV}^{-1}$. The value of the constant $f_2$
can be extracted from the proton-neutron electromagnetic ~mass
~difference~\cite{Reports}: $e^2f_2=(-0.76\pm 0.3)~{\rm MeV}$. The 
determination of the constant $f_1$ from data is however, problematic. For
this reason, in our analysis we have used order-of-magnitude estimate for 
this constant: $-|f_2|\leq f_1\leq |f_2|$. With these values of the low-energy
constants, we obtain the isospin-breaking correction to be
$\delta_\epsilon=(-4.7\pm 2.0)\cdot 10^{-2}$, again in striking disagreement
from the potential model prediction~\cite{PSI1}
$\delta_\epsilon=(-2.1\pm 0.5)\cdot 10^{-2}$. It remains
to be seen, how the $O(p^2)$ results are altered by the loop corrections in
ChPT~\cite{Ivanov}.   

Given a systematic discrepancy of the potential model predictions with the
results of calculations based on ChPT, it is natural to seek a derivation
of the potentials that are used in the potential model, on the basis of ChPT.
In a slightly more restricted context, one may ask, how the isospin-breaking
part of the short-range ``strong'' potential is obtained from ChPT, when
the isospin-symmetric part is already known to fit well ChPT predictions
(we recall that the iso\-spin-breaking part is assumed to vanish identically in
existing potential models~\cite{PSI1,Rasche}).
It is widely presumed that the potential constructed from the field theory
will be necessarily singular in the position space and will require some
kind of regularization. Based on a simple solvable model, we shall however
demonstrate that this is not the case: almost any well-behaved 
short-range potential, including those that were used in
Refs.~\cite{PSI1,Rasche}, can be generalized to include properly the full
content of isospin-breaking effects in ChPT.

The key to the solution of the problem given above, lies in the
universality conjecture. This conjecture - completely in spirit of the
low-energy effective Lagrangian approach to bound systems - states that
the bound-state energies in the field theory, and in the potential model are
the same at the first order in isospin breaking, provided the quantities 
${\cal A}_{\pi N}$ calculated in these two theories, coincide. We shall
ensure the universality for the case of a simple model where the
interaction Hamiltonian is given by a sum of Coulomb
and short-range rank-1 separable interactions
\begin{eqnarray}\label{potential}
&&U({\bf p},{\bf k})={\bf C}+{\bf V}=-\,\frac{4\pi\alpha}{|{\bf p}-{\bf k}|^2}
+\lambda v({\bf p})v({\bf k})\, ,
\nonumber\\[2mm]
&&v({\bf q})=\frac{\beta^2}{\beta^2+{\bf q}^2}\, ,
\end{eqnarray}
where $\beta$ denotes the cutoff mass. The generalization to the case of
generic potentials, multichannel case, inclusion of relativistic effects, is
in progress and will be reported elsewhere~\cite{Lipartia}.

With a given interaction potential, one may evaluate the energy-level shift
of the ground state of the bound system. The equation for the position
of the bound-state pole in the (complex) energy plane is given
by~\cite{Bern1}
\begin{eqnarray}\label{pole-position}
&&z-E_0-\langle\Psi_0|\tau(z)|\Psi_0\rangle=0\, ,
\\[2mm]
&&\tau(z)={\bf V}+
{\bf V}(z-{\bf H}_{\rm 0})^{-1}(1-|\Psi_0\rangle\langle\Psi_0|)\tau(z)\, .
\nonumber
\end{eqnarray}
It is straightforward to solve this equation in the perturbation theory. 
The solution at the first non-leading order in $\alpha$ reads
\begin{eqnarray}\label{deltaE}
\Delta E&=&\Psi_0^2\,\tau_0\,\biggl(1-\frac{4\alpha\mu}{\beta}
\nonumber\\[2mm]
&+&\frac{\mu^2\alpha}{\pi}\,\tau_0\,
[\ln\alpha-1+\ln\frac{4\mu}{\beta}]\biggr)\, ,
\nonumber\\[2mm]
\tau_0&=&\lambda\biggl(1+\frac{\mu\lambda\beta}{4\pi}\biggr)^{-1}\, ,
\end{eqnarray}
and $\mu$ stands for the reduced mass.

Next, we calculate the regular part of the elastic scattering amplitude
at threshold, with the normalization according to Eq.~(\ref{threshold-piN})
\begin{eqnarray}\label{curlyA}
{\cal A}=-\frac{\mu}{2\pi}\, \tau_0\biggl(1-\frac{4\alpha\mu}{\beta}\biggr)
-\frac{\alpha\mu^3}{2\pi^2}\, \tau_0^2\ln\frac{4\mu}{\beta}
\end{eqnarray}

From Eqs.~(\ref{deltaE}) and (\ref{curlyA}) one immediately obtains that
the energy level shift in the potential model is given again by
Eq.~(\ref{general-piN}) - that is, the universality holds in that particular
case, considered here.

Based on the universality conjecture, we can provide a constructive algorithm
for the derivation of the isospin-breaking part of the short-range
potential ${\bf V}$ from ChPT. The amplitude at threshold in the latter
is generally given by ${\cal A}={\cal A}_0+{\cal A}_1+\cdots$, where
${\cal A}_{0(1)}$ denote the isospin-conserving(breaking) parts of the 
amplitude, and ellipses stand for higher-order terms in isospin breaking.
In order to ensure the inclusion of the full content of isospin-symmetry
breaking in ChPT into the potential model, it thus suffices to match the
amplitude ${\cal A}$ in both theories. The problem evidently has too much
freedom, and we can choose one parameter in the potential - say, the coupling
constant $\lambda$ - to obey the matching condition for the amplitude.
Writing $\lambda=\lambda_0+\lambda_1+\cdots$, the matching condition yields
\begin{eqnarray}\label{matching}
\lambda_0&=&
-\frac{2\pi}{\mu}\,\,\frac{{\cal A}_0}{1+\frac{\beta}{2}{\cal A}_0}\, ,
\\[2mm]
\lambda_1&=&-\frac{2\pi}{\mu}\,\,\frac{{\cal A}_1+\frac{4\alpha\mu}{\beta}\,
{\cal A}_0+2\mu\alpha{\cal A}_0^2\ln\frac{4\mu}{\beta}}
{(1+\frac{\beta}{2}{\cal A}_0)^2}\, .
\nonumber
\end{eqnarray}

The matching condition~(\ref{matching}) solves our problem completely -
the bound-state energies calculated with the use of the ``corrected''
potential coincide, by definition, with those calculated on the basis of
ChPT. We hope that - after the suitable generalization - the approach based
on the universality might be also useful for the analysis of $\pi N$ 
scattering data near threshold, in what concerns the study of the 
isospin-breaking effects in the $\pi N$ amplitude.

{\it Acknowledgments}. 
This work was supported in part by the Swiss National Science
Foundation, and by TMR, BBW-Contract No. 97.0131  and  EC-Contract
~No. ERBFMRX-CT980169 (EURODA$\Phi$NE).


\begin{thebibliography}{999}
\bibitem{Lepage} 
W.E.~Caswell and G.P.~Lepage, \plb{167}{1986}{437}.

\bibitem{DIRAC}
B.~Adeva {\it et al.}, CERN proposal CERN/SPSLC 95-1 (1995). 

\bibitem{ChPT} J.~Gasser and H.~Leutwyler, \plb{125}{1983}{325};
J.~Bijnens, G.~Colangelo, G.~Ecker, J.~Gasser,  
and M.E.~Sainio, \plb{374}{1996}{210}.

\bibitem{Stern} M.~Knecht, B.~Moussallam, J.~Stern, and N.H.~Fuchs, 
\npb{457}{1995}{513}; \npb{471}{1996}{445}.

\bibitem{PSI1}
D.~Sigg, A.~Badertscher, P.F.A.~Goudsmit, H.J.~Leisi, and G.C.~Oades,
\npa{609}{1996}{310}. 

\bibitem{PSI2}
H.-Ch. Schr\"oder {\it et al}., \plb{469}{1999}{25}. 

\bibitem{DEAR}
The DEAR collaboration (S. Bianco et al.), {\it The DEAR case}, preprint 
LNF-98/039(P).

\bibitem{KEK}
M.~Iwasaki {\it et al.}, \prl{78}{1997}{3067};
\npa{639}{1998}{501}.

\bibitem{Rasche} 
U.~Moor, G.~Rasche, and W.S.~Woolcock, 
\npa{587}{1995}{747};
A.~Gashi, G.C.~Oades, G.~Rasche, and W.S.~Woolcock, 
\npa{628}{1998}{101}.

\bibitem{Bern1}
A.~Gall, J.~Gasser, V.E.~Lyubovitskij, and A.~Rusetsky, 
\plb{462}{1999}{335}.

\bibitem{Bern2}
J.~Gasser, V.E.~Lyubovitskij and A.~Rusetsky,
\plb{471}{1999}{244}.

\bibitem{Sazdjian}
H. Jallouli and H. Sazdjian, \prd{58}{1998}{014011};
H.~Sazdjian, Preprint \hepph{9809425}. 

\bibitem{Dubna}
V.E. Lyubovitskij and A.G. Rusetsky, \plb{389}{1996}{181};
V.E. Lyubovitskij, E.Z. Lipartia, and A.G. Rusetsky, 
{\it JETP Lett.} {\bf 66} (1997) 783;
M.A. Ivanov, V.E. Lyubovitskij, E.Z. Lipartia, and A.G. Rusetsky, 
\prd{58}{1998}{094024}. 

\bibitem{Bern3} 
J.~Gasser, V.E.~Lyubovitskij, and A.~Rusetsky, in preparation.

\bibitem{Lagrangian} 
U.-G.~Mei\ss ner {\it et al}, \plb{419}{1998}{403};
T. Becher and H.~Leutwyler, {\it Eur. Phys. J.} {\bf C 9} (1999) 643.

\bibitem{private}
T. Becher and H. Leutwyler, private communication.

\bibitem{Reports}
J. Gasser and H. Leutwyler, {\it Phys. Rep.} {\bf 87} (1982) 77.

\bibitem{Ivanov}
J. Gasser, M.A. Ivanov, V.E. Lyubovitskij, and A. Rusetsky, work in progress.

\bibitem{Lipartia}
J. Gasser, E. Lipartia, V.E. Lyubovitskij, and A. Rusetsky, work in progress.

\end{thebibliography}
\end{document}